\documentclass[12pt,preprint]{aastex}

\shorttitle{Slow Expansion of RX J0852-4622 NW Rim}
\shortauthors{Katsuda, Tsunemi, \& Mori}

\begin{document}

\title{The Slow X-Ray Expansion of the Northwestern Rim of the Supernova
  Remnant RX J0852.0-4622}

\author{S. Katsuda\altaffilmark{1}, H. Tsunemi\altaffilmark{1}, and
  K. Mori\altaffilmark{2}
}

\altaffiltext{1}{Department of Earth and Space Science, Graduate School
of Science, Osaka University,\\ 1-1 Machikaneyama, Toyonaka, Osaka,
560-0043, Japan; katsuda@ess.sci.osaka-u.ac.jp}

\altaffiltext{2}{Department of Applied Physics, Faculty of Engineering,
University of Miyazaki, 1-1 Gakuen Kibana-dai Nishi, Miyazaki, 889-2192,
Japan}

\begin{abstract}

The detection of radioactive decay line of $^{44}$Ti provides a unique 
evidence that the $\gamma$-ray source is a young ($<$ 1,000 yr) supernova
remnant because of its short lifetime of $\sim$90 yr. Only two Galactic
remnants, Cassiopeia A and RX J0852.0-4622, are hitherto reported to
be the $^{44}$Ti line emitter, although the detection from the latter
has been debated. Here we report on an expansion measurement of the
northwestern rim of RX J0852.0-4622 obtained with X-ray observations
separated by 6.5 yr. The expansion rate is derived to be
0.023$\pm$0.006 \% that is about five times
lower than those of young historical remnants. Such a slow expansion
suggests that RX J0852.0-4622 is not a young remnant as has been
expected. We estimate the age of 1,700-4,300 yr of this remnant
depending on its evolutionary stage.  Assuming a high shock speed of
$\sim$3000\,km\,sec$^{-1}$, which is suggested by the detection of
non-thermal X-ray radiation, the distance of $\sim$750 pc to this
remnant is also derived. 

\end{abstract}
\keywords{ISM: individual (RX J0852.0-4622) --- shock waves ---
  supernova remnants --- X-rays: ISM}

\section{Introduction}

Ten years have passed after the discovery of the supernova remnant
(SNR) RX J0852.0-4622.  It was uncovered in the southeastern corner of
the Vela SNR from the high energy band (above 1.3 keV) image obtained
by the {\it ROSAT} all-sky survey (Aschenbach 1998). The shape is
nearly a perfect circle with a large angular radius, $\Theta$, of
$\sim$60 arcmin. The discovery of this remnant was accompanied by a
report of the {\it COMPTEL} 
detection of $\gamma$-ray line emission from $^{44}$Ti (Iyudin et al.\
1998), suggesting that this new remnant is a young and nearby because
of the very short lifetime ($\sim$90 yr) of $^{44}$Ti.  Combining it with
X-ray and $\gamma$-ray data, the age and the distance of this remnant are
estimated to be $\sim$680 yr and $\sim$200 pc, respectively
(Aschenbach et al.\ 1999). So far, most of the
follow-up works support the young (less than 1000 yr) age of this
remnant (Tsunemi et al.\ 2000; Iyudin et al.\ 2005; Bamba et al.\
2005).  However, the detection of $^{44}$Ti line itself has been 
debated; independent reanalysis of the {\it COMPTEL} data finds that the
detection of this remnant as a $^{44}$Ti source is only significant at the
2-4 sigma level (Sch$\ddot{\mathrm o}$nfelder et al.\ 2000).  
In addition, Slane et al.\ (2001) questioned the close distance to the
remnant, based on their analysis of {\it ASCA} data; a larger column density
for this remnant than that for the Vela SNR indicated that RX
J0852.0-4622 was at a larger distance of 1--2\,kpc.  The age
and the distance are quite curious in the light of the other
exceptional natures of this remnant, namely predominance of
non-thermal X-ray radiation (e.g., Tsunemi et al.\ 2000, Slane et al.\ 
2001), existance of an enigmatic central-compact object (e.g., Aschenbach
1998; Kargaltsev et al.\ 2002), and detection of the TeV $\gamma$-ray
emission (Aharonian et al.\ 2007).  

If RX J0852.0-4622 is really as young as 680 yr, its large apparent
radius of about 60$^{\prime}$ indicates that the proper motion of the
shock front would be as fast as 5$^{\prime\prime}$.3
($\Theta$/60$^{\prime}$)($t$/680 yr)$^{-1}$ per year assuming 
the free expansion of the shock.  The expected motion is large enough to be
measured with current X-ray observatories. We perform the first
measurement of the shock expansion rate of the northwestern (NW) rim
of RX J0852.0-4622 from XMM-Newton observations taken over 6.5 yr. 

\section{Observations and Data Reduction}

The NW rim of RX J0852.0-4622 has been observed several times with
{\it XMM-Newton}.  Four observations performed in 2001, 2003, 2005,
and 2007 are analyzed here.  The information of the four
observations is summarized in Table.~\ref{tab:obs}.  The time
difference between the first (2001) and the last (2007) one is exactly
6.5 yr. In this interval, the shift of the undecelerated shock front is
expected to be $\sim$35$^{\prime\prime}$.  We note that the difference
between the Medium filter (for 2001 and 2003 observations) and the
Thin1 filter (for 2005 and 2007 observations) has negligible effects
in the following analysis, in which we only use high energy events of 
1.5--8\,keV band.  
All the raw data were processed with version 7.1.0 of the XMM Science
Analysis Software (XMMSAS).  We concentrate on the data only taken by
European Photon Imaging Camera (EPIC) MOS1 and MOS2 detectors, since
the spatial resolution of the EPIC MOS is slightly better than that of
the EPIC pn detector.  We select X-ray events corresponding to patterns
0--12.  We further clean the data by rejecting high background (BG)
intervals.  After the filtering, the data were vignetting-corrected using
the XMMSAS task {\tt evigweight}.  We further need to subtract the
cosmic X-ray BG and the Cosmic Ray (CR) induced BG at energies above
typically 1.5\,keV (Arnaud et al.\ 2001).  To this end, we subtract
the data set accumulated from blank sky observations prepared by Read
\& Ponman (2003). 

\section{Analysis}

Figure~\ref{fig:image} {\it left} shows the BG-subtracted {\it
XMM-Newton} hard band (1.5--8 keV) image of the NW rim of RX
J0852.0-4622. We can
clearly see sharp filamentary structures whose emission is dominated
by non-thermal emission (Tsunemi et al.\ 2000; Slane et al.\ 2001).
These structures mark the current locations of the shock fronts of the
NW rim of RX J0852.0-4622.  There is a significant X-ray contamination 
from the Vela SNR along the line of sight.  However, the emission is
negligible in this hard energy band, since it is believed to be
dominated by thermal emission with the electron temperature of below
0.3\,keV (e.g., Iyudin et al.\ 2005).   

According to the {\it XMM-Newton} calibration status report (Kirsch
2007), the absolute astrometric accuracy, i.e., the precision with which
astronomical coordinates can be assigned to source images in the EPIC
focal plane, is less than 2$^{\prime\prime}$.  This value is small
enough to detect the expected shift of the undecelerated shock front,
$\sim$35$^{\prime\prime}$ per 6.5 yr.  We check whether the same
position accuracy is achieved in our data.  Applying an XMMSAS tool
{\tt edetect\_chain}, we determine positions of three point sources
(P1--P3) indicated in Fig.~\ref{fig:image} {\it left}.  Naval
Observatory Merged Astrometric 
Dataset (NOMAD) catalog (Zacharias et al.\ 2005) identifies several
stars as possible optical counterparts for three point sources.
We check that the proper motion of the possible optical counter parts
themselves are all less than $\sim$1$^{\prime\prime}$ per 6.5
yr.  Therefore, we expect to detect the X-ray point sources within
circles with radius of $\sim$1$^{\prime\prime}$ in all the four {\it
XMM-Newton} observations.  In fact, we find that all the positions
determined in our four X-ray data sets (MOS1 and MOS2 separately) are
well within 2$^{\prime\prime}$ circles around their mean positions.
Therefore, without extra corrections of the coordinates for our
data, the absolute astrometric accuracy is achieved to be less than
2$^{\prime\prime}$.  In the following analysis, we take account of 
2$^{\prime\prime}$-error for the position accuracy as the conservative
systematic uncertainty.  

The difference of the 2001 and 2007 images of 1.5--8\,keV band is
shown in Figure~\ref{fig:image} {\it right}.  We can clearly see a 
black (negative) narrow line running from the northeast (NE) to the
southwest (SW) as a sign of the expansion of the shock front in 6.5
yr.  Other black or white lines are due to artificial effects such as
bad columns or gaps of CCD chips. 

Next, we quantitatively measure the shift of the X-ray filament based
on one-dimensional profiles across the filament.  We select a northern
portion of the narrow NW filament as shown in Fig.~\ref{fig:image}
{\it left} and {\it right}, since we find few bad columns there.  We slice the
area into 2$^{\prime\prime}$-spaced regions parallel to the filament.
The BG-subtracted radial profiles for the four observations are
plotted in Fig.~\ref{fig:profile}.  In the figure, we clearly see that
the peak position of the filament at around
R$\sim$115$^{\prime\prime}$ in 2001 is shifted to
R$\sim$120$^{\prime\prime}$ in 2007.  

In order to quantitatively measure shifts, we also apply the method
of calculating $\chi^2$ probability that two observed profiles from
different epochs have the same shapes.  Let $l$ be the shift parameter,
and $\theta$ be the angular distance perpendicular to the shock
front.  Then we calculate the $\chi^2$ value as

\[\chi^2(l) = \sum_{\theta} \frac{[x_1(\theta+l)-x_2(\theta)]^2}{\sigma_{1}^2(\theta+l)+\sigma_{2}^2(\theta)}.
\]

Here, $x_1(\theta)$ and $x_2(\theta)$ represent the observed count rates in
angular distance $\theta$ at epochs 1 and 2, and $\sigma_1(\theta)$ and
$\sigma_2(\theta)$ represent the uncertainties at each bin.  The minimum
for $\chi^2(l)$, $\chi^2_\mathrm{min}$, is around 39, since we sum 40
bins roughly around the shock front.  We shift the radial profile of
2001 and compare the shifted profile with the profiles of 2003, 2005,
and 2007.  We examine $l$ = 0$^{\prime\prime}$, 2$^{\prime\prime}$,
4$^{\prime\prime}$, 6$^{\prime\prime}$, 8$^{\prime\prime}$, and
10$^{\prime\prime}$.  The $\chi^2(l)$-values as a function of $l$ are
shown in Fig.~\ref{fig:chi2}.  In 
the figure, three kinds of data points with rectangular, circular, or
triangular marks are, respectively, responsible for three cases in which
we focus on different two epochs, i.e., 2001-2003, 2001-2005, and
2001-2007.  We find that $\chi^2_\mathrm{min}$ occurs
at $l$ = 0$^{\prime\prime}$, 4$^{\prime\prime}$, and
6$^{\prime\prime}$ for each two-epoch of 2001-2003, 2001-2005, and 
2001-2007, respectively.  Using the criteria of $\chi^2_\mathrm{
  min} + 2.7$, we can determine the values of the shifts less than
2$^{\prime\prime}$, at 90\% confidence level in all the cases.  With
taking into account the systematic uncertainty, we derive the shifts
of the X-ray filament in 2003, 2005, and 2007 from 2001 to be
0$\pm$2$^{\prime\prime}$, 4$\pm$2$^{\prime\prime}$, and
6$\pm$2$^{\prime\prime}$, respectively.  These values lead us to
calculate the proper motion to be 0.84$\pm$0.23 arcsec yr$^{-1}$,
assuming the constant velocity of the shock front in the four
observations.  Then, the expansion rate is calculated to be
0.023$\pm$0.006 \% yr$^{-1}$.  

\subsection{Discussion and Conclusion}

Assuming that the shock speed has not decelerated (i.e., the free
expansion of the shock), a rigid upper limit of the age for RX
J0852.0-4622 is calculated to be 4300$\pm$1200
($\Theta$/60$^{\prime}$)($\mu$/0$^{\prime\prime}$.84 yr$^{-1}$)$^{-1}$
yr, where $\mu$ is the proper motion at present.
Table.~\ref{tab:param} summarizes expansion rates, ages,  
velocities of the forward shock $v$, and expansion indices $m$ for
several SNRs, where $m$ is defined as the ratio between the current
shock velocity and the mean shock velocity that is gradually decreases
from 1 (free expansion phase) to $\sim$0 (disappearance phase) through
0.4 (Sedov phase), 0.3 (radiative cooling phase), and 0.25
(pressure-driven snowplough phase) (e.g., Woltjer 1972). 
We notice that the expansion rate of the NW rim of RX J0852.0-4622 is
about five times lower than those in the Cas A SNR ($\sim$320 yr),
Kepler's SNR ($\sim$400 yr), and Tycho's SNR ($\sim$430 yr), whereas
it is comparable to that determined for the NW filaments in SN1006
($\sim$1000 yr).  Assuming the young currently best-estimated age of
680 yrs, we derive the value of $m$ for RX J0852.0-4622 to be
0.16$\pm$0.04 ($\mu$/0$^{\prime\prime}$.84 yr$^{-1}$)($t$/680 
yr)($\Theta$/60$^{\prime}$)$^{-1}$.  This value is similar to that in
the Cygnus Loop which is a 
representative remnant in the further later phase than the Sedov
phase.  Therefore, the young age of 680 yr 
is apparently inconsistent with the value derived.  In this context,
it is natural to consider that RX J0852.0-4622 is not a young SNR but
a middle-aged one.  In the upper limit of the age of $\sim$4,300 yr,
it is reasonable to consider that this remnant is in no later phase
than the Sedov phase ($m$=0.4).  Then, we estimate the age of this
remnant to be $\sim$1,700 ($m$/0.4) yr. 

Within the currently accepted theory of diffusive shock acceleration,
non-thermal synchrotron radiation in the X-ray band requires a high
shock speed, $v\sim$3,000\,km\,sec$^{-1}$ (Uchiyama et al.\ 2003;
Zirakashvili \& Aharonian 2007).  Such a high shock speed is actually
derived about other four SNRs with non-thermal X-ray radiation (see
Table.~2).  With the derived proper motion of the shock front, the
distance to the remnant, D, is obtained as D$\sim$750
($v$/3000\,km\,s$^{-1}$)($\mu$/0$^{\prime\prime}$.84 yr$^{-1}$)$^{-1}$  
pc.  Although the distance to this remnant is also highly uncertain,
our low expansion rate measured supports a relatively distant value
within the suggested range of $\sim$0.2 to $\sim$1\,kpc (Iyudin et
al.\ 1998; Ashchenbach et al.\ 1999).  Since our distance estimation
relies on only an assumption of the shock speed, which has theoretical
and observational bases, we belive that our value is the most reliable
one so far, and thus allow more conclusive discussion about the nature
of this remnant, e.g., the origin of the TeV $\gamma$-ray emission 
(Aharonian et al.\ 2007).

We have estimated the age of this remnant to be 1,700-4,300 yr,
which is at least 2.5 times larger than the previously estimated age
of 680 yr.  The new age determined here critically affects the
estimation of the initial amount of $^{44}$Ti; it must be about 10,000
times of that assumed in the paper which reported the {\it COMPTEL}
detection of the $\gamma$-ray line at 1.157 MeV from $^{44}$Ti
associated with this remnant (Iyudin et al.\ 1998).  In addition, we
should note that the distance estimated here is about four times that
derived in the paper, which requires about 16 times larger initial
amount of the $^{44}$Ti than that assumed in the paper.  As a result,
the initial mass of $^{44}$Ti is estimated to be a few solar masses,
using the $\gamma$-ray line flux derived by {\it COMPTEL}. Such a large
amount of $^{44}$Ti is far from reality; at least four orders of
magnitude larger than that expected in nucleosynthesis models
(Theilemann et al.\ 1996; Rauscher et al.\ 2002).  Therefore, it is
very likely that the {\it COMPTEL} detection of the $\gamma$-ray line at
1.157 MeV around RX J0852.0-4622 has no relations to the decay line
from $^{44}$Ti associated with this remnant. 

\begin{figure}
\includegraphics[angle=0,scale=0.8]{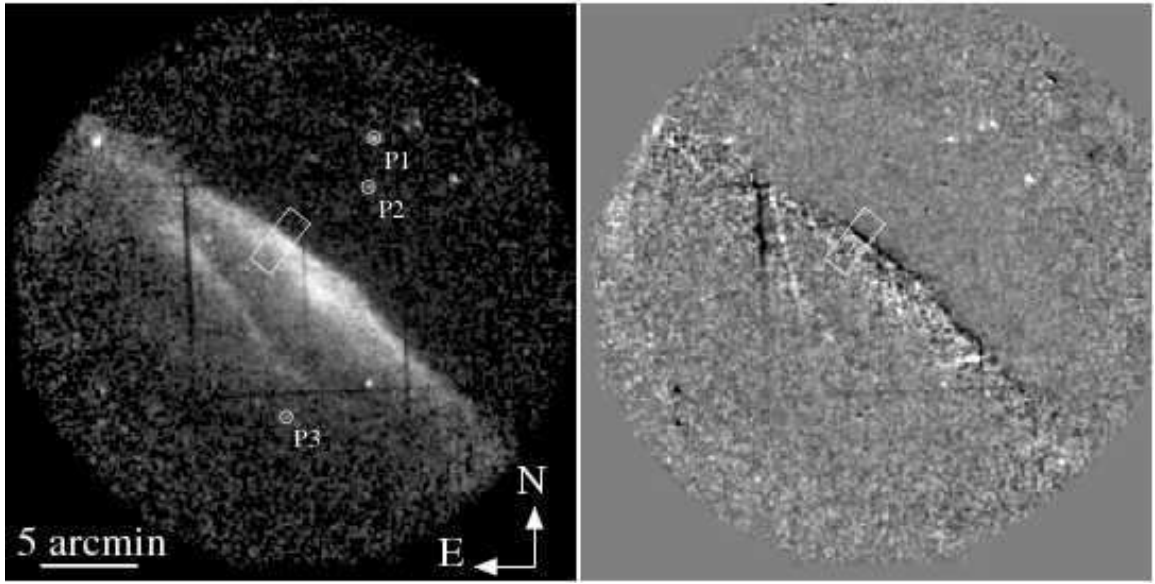}\hspace{1cm}
\caption{{\it Left}: {\it XMM-Newton} 1.5--8\,keV band image obtained
  in 2001.  The image is binned by 5$^{\prime\prime}$ and has been
  smoothed by Gaussian kernel of $\sigma = 10^{\prime\prime}$.  The
  intensity scale is square root.  Point source positions which we use
  to examine the astrometric accuracy are indicated as P1, P2, and P3.
  We investigate the radial profile of the X-ray filament in the
  rectangular area.  {\it Right}: Same image but  
  subtracted one obtained in 2007.  The intensity is linearly scaled
  from $-1.5\times10^{-4}$ to $+1.5\times10^{-4}$
  counts\,sec$^{-1}$\,pixel$^{-1}$. 
} 
\label{fig:image}
\end{figure}

\begin{figure}
\includegraphics[angle=0,scale=0.65]{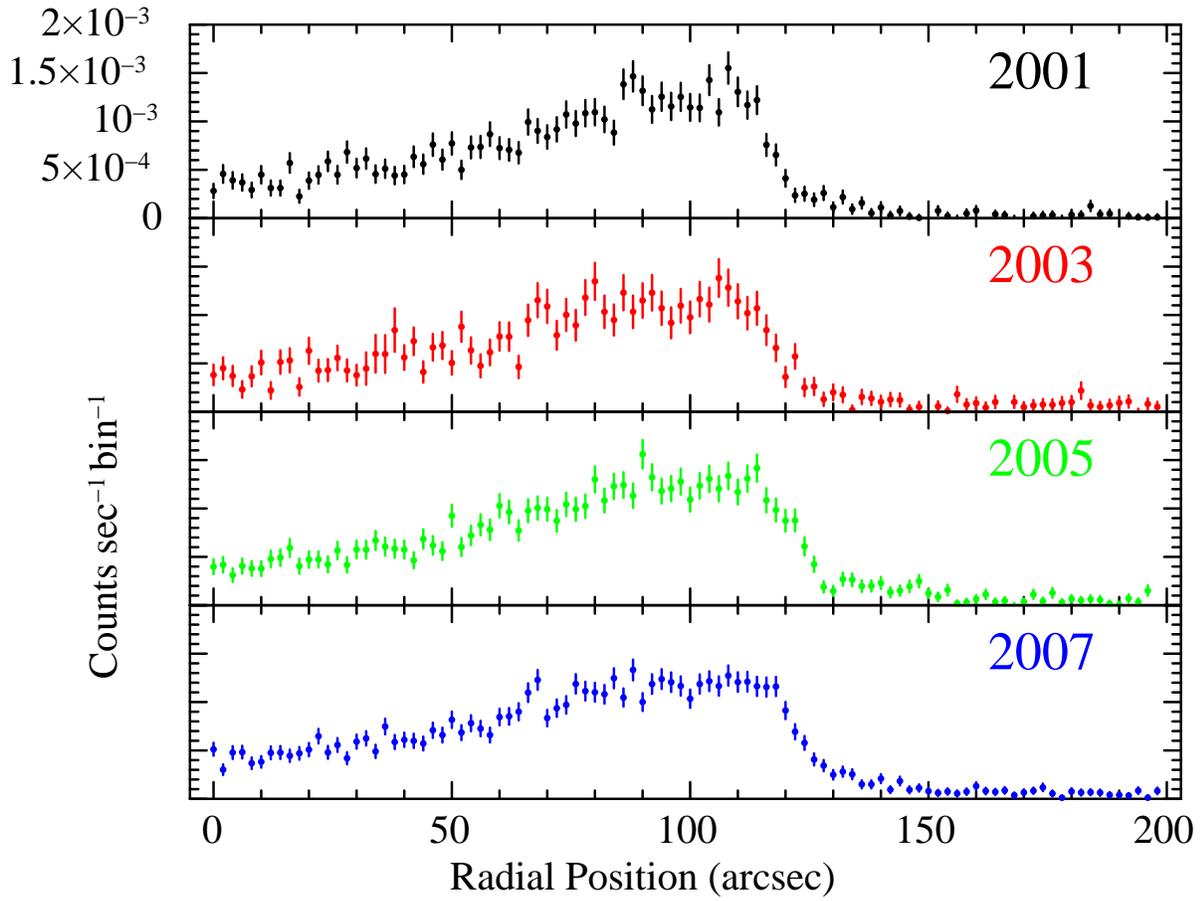}
\caption{Radial profiles at each epoch in the 1.5--8\,keV band, binned
  with a 2$^{\prime\prime}$ scale.  Four profiles from top to bottom
  are responsible for the one in 2001, 2003, 2005, and 2007.}  
\label{fig:profile}
\end{figure}

\begin{figure}
\includegraphics[angle=0,scale=0.65]{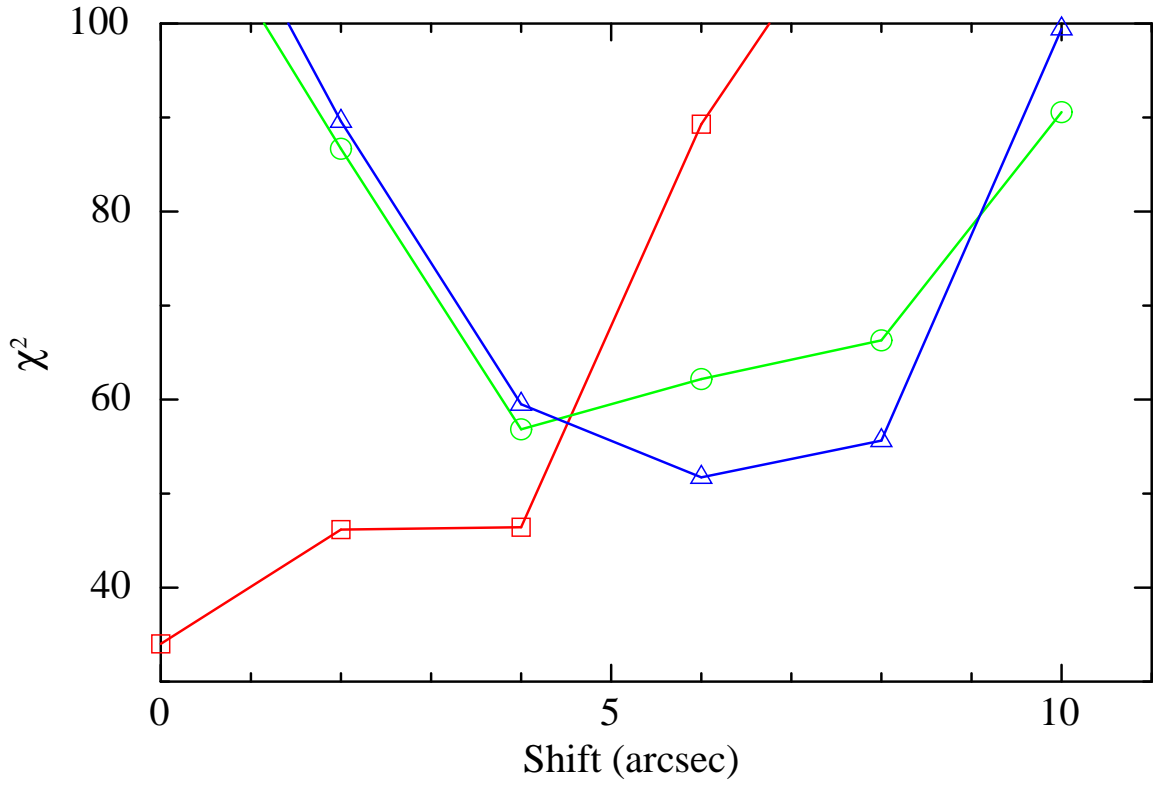}
\caption{$\chi^2$ distribution as a function of the shift parameter,
  $l$.  Data points with rectangular, circular, and triangular marks 
  are responsible 
  for three cases in which we focus on different two epochs, i.e.,
  2001-2003, 2001-2005, and 2001-2007, respectively.} 
\label{fig:chi2}
\end{figure}

Finally, we note a possiblity to explain the current low expansion
rate without any modifications of the age of this remnant: the forward
shock recently encountered a dense interstellar medium (or a cloud) in the
NW rim and was rapidly decelerated.  If such an interaction really
occurred, it would be a very recent event so that it did not modify
the nearly perfect circular shape of this remnant. Also, a rapid and
strong deceleration of the shock would cause a reflection shock near
the forward shock.  A filamentary structure somewhat distant (about
2$^{\prime}$) behind the forward shock (see, Fig.~\ref{fig:image} {\it
  left}) is suggested to be a hint of the reflection shock.  However,
based on the radial profile of the filament obtained 
by {\it Chandra} (see, Fig.~2 in Bamba et al.\ 2005), this suggestion
is unlikely and there is no other implications of the reflection
shock.  Although we can not still completely exclude the possibility
of such a rapid deceleration, we think that more straightforward
interpretation of the current slow expansion is simply because this
remnant is relatively old.  Further expansion measurements of the
other rim of this remnant will clearly reveal which hypothesis is at
work.  

\acknowledgments

This work is partly supported by a Grant-in-Aid for Scientific Research
by the Ministry of Education, Culture, Sports, Science and Technology
(16002004). The work of K.M.\ is partially supported by the
Grant-in-Aid for Young Scientists (B) of the MEXT (No.\ 18740108).
S.K.\ is supported by JSPS Research Fellowship for Young Scientists.

\clearpage

\begin{deluxetable}{ccccccrrrrcrl}
\tabletypesize{\scriptsize}

\tablecaption{XMM-Newton observations}
\tablewidth{0pt}
\tablehead{
\colhead{Obs.\ ID} &\colhead{Camera} &\colhead{Instrument Mode}  &
\colhead{Filter} & \colhead{Obs.\ Date}& \colhead{Good Time Interval}
}
\startdata
0112870301 & MOS1/2 & PrimeFullWindow & Medium & 2001-04-25 & 31.3\,ksec\\
0159760101 & MOS1/2 & PrimeFullWindow & Medium & 2003-06-22 & 19.5\,ksec\\
0159760301 & MOS1/2 & PrimeFullWindow & Thin1 & 2005-11-01 & 38.0\,ksec\\
0412990201 & MOS1/2 & PrimeFullWindow & Thin1 & 2007-10-24 & 62.6\,ksec\\
\enddata

\label{tab:obs}
\end{deluxetable}

\begin{deluxetable}{lccccc}
\tabletypesize{\scriptsize}

\tablecaption{Physical parameters of several SNRs}
\tablewidth{0pt}
\tablehead{
\colhead{SNR Name} &\colhead{Expansion Rate (\%)} &\colhead{Age (yr)
} &\colhead{Velocity (km\,sec$^{-1}$)} &\colhead{Expansion Index}
&\colhead{References}  
}
\startdata
Kepler (mean)\dotfill& $\sim$0.24\% &390 &4,800 (D/5\,kpc)& $\sim$0.9 & 1\\
Cas~A (mean)\dotfill& $\sim$0.20\%&350 &3,200 (D/3.4\,kpc)& $\sim$0.7 & 2, 3\\
Tycho (mean)\dotfill& $\sim$0.12\%&430 &3,300 (D/2.3\,kpc)& $\sim$0.54 & 4\\
SN1006 (NW filament)\dotfill& $\sim$0.03\%&1,000 &3,100 (D/2.2\,kpc)& $\sim$0.34 & 5\\
Cygnus Loop (NE filament)\dotfill& $\sim$0.003\% &10,000 &180 (D/0.54\,kpc)& $\sim$0.17& 6\\
\enddata

\tablecomments{1: Hughes (1999), 2: Koralesky et al.\ (1998), Vink et al.\ 
  (1998), 4: Hughes (2000), 5: Winkler et al.\ (2003), 6: Blair et
  al.\ (2005)}
\label{tab:param}
\end{deluxetable}

\end{document}